\documentstyle[12pt]{article}
\begin{document}
\title
{Decay Process  for Three - Species Reaction - Diffusion System}
\author
{ Kyungsik Kim \thanks{E-mail: kskim@dolphin.pknu.ac.kr}
and K. H. Chang\\
Department of Physics, Pukyong National University, \\
Pusan 608-737, Korea
\and
Y.S. Kong\\ 
School of Ocean Engineering, Pukyong National University,\\
Pusan 608-737, Korea}
\date{ }
\maketitle\abstract
{    We propose the deterministic rate equation of three-species in the reaction - 
diffusion system. For this case, our purpose is to carry out 
the decay process in our three-species reaction-diffusion model of the form $A+B+C\rightarrow D$. 
The the particle density and the global reaction rate are also shown 
analytically and numerically on a two - dimensional square lattice 
with the periodic boundary conditions. 
Especially, the crossover of the global reaction rate is discussed 
in both early-time and long-time regimes.}
%
%
%
%
\newpage
%
\indent 
Recently, there has been considerable interest in 
the front propagation under the initial segregation of two - species reactants
in early - time  reaction - diffusion process.
It is furthermore well - known that
the binary reaction has mainly been investigated  
on the process of ternary reactions$^{1)}$
and many phenomena 
in nature.$^{2-6)}$
The  early - time behavior for reaction - diffusion system have been argued by 
Taitelbaum et al,$^{7)}$
and it has precisely been shown 
that both the global reaction rate and
the reaction front increase as a function of $t$$^{1/2}$ at very early times.
On the other hand, Cornell et al$^{8,9)}$
have recently treated with the diffusion - limited reaction $nA+mB$$\rightarrow$$C$ 
for both homogeneous and inhomogeneous conditions
under initially separated reactants. They have found that the global reaction rate 
decreases as  $t^{-1/2}$ at long -  time regime, independent of $n$ and $m$, and 
discussed that the upper critical dimension is $d=2$ for the reaction - diffusion
process. 
Very recently, Yen et al$^{10)}$ also have studied for the asymptotic 
early - time scaling in the ternary $A+2B$$\rightarrow$$C$ reaction - diffusion process 
with initially separated reactants. 
To our knowledge,
it is of fundamental importance to 
measure extensively
more complicated reaction such as multi - components reaction - diffusion process,
even though the three - species reaction may occur rarely in the fields of  physics,
chemistry, and biology.
\\
\indent
In the present study, we present the decay process in a reaction-diffusion system with
three - species.
By a simple perturbation expansion, we derive analytically the particle density and
the global reaction rate before the crossover  in the
reaction - diffusion system  of $A + B + C \rightarrow D $.
The scaling exponent for the global reaction rate is also confirmed numerically
before and after the crossover via Monte Carlo simulations.\\
%
%
\indent 
First of all, let us denote that $ \rho_A (x,t)$,  $\rho_B (x,t)$, and  $\rho_C (x,t)$ are the particle
densities for three - species $A$, $B$, and $C$ 
existing at a position $x$ at time $t$. 
As we assume that three - species particles are distributed separately on the axis $x$,
the special initial condition for these particle densities is as follows :
$\rho_A(x,0)  =  A_0 H(x)$, $\rho_B (x,0)  =  B_0 H(x)$,
and $\rho_C(x,0)  =  C_0 [1-H(x)]$.
$A_0$, $B_0$, and $C_0$ are the initial particle densities, and
$H(x)$ is the Heaviside step function.
We assume that both $A_0$ and $B_0$ have a uniform mixing state on $0 < x < \infty$.
Since the mean field approximation may be applied to three - species system,  
$A + B + C \rightarrow D $, 
the deterministic rate equation for $\rho_A (x,t)$ can be expressed in terms of 
\\
\begin{equation}
\frac{\partial}{\partial t} \rho_A (x,t) =  D_A {\nabla}^2  \rho_A (t) - 
K \rho_A (x,t)  \rho_B (x,t)  \rho_C (x,t),
\label{eq:b2}
\end{equation}                             
\\
where  $D_A$ is the diffusion constant for one - species $A$.
In the above equation $K$
is the reaction rate which can be considered as small value in reaction - limited reaction.\\
%
\indent  
In particular, when
we introduce the dimensionless particle densities $\alpha(x,t)$, $\beta(x,t)$,
and $\gamma(x,t)$ from the relations of
$\rho_A(x,t)  =  A_0 \alpha(x,t)$, $ \rho_B (x,t)  =  B_0 \beta(x,t)$,                
and $\rho_C(x,t)  =  C_0 \gamma(x,t)$,                    
eq.(~\ref{eq:b2}) for $\alpha(x,t)$ can be rewritten as 
\\
\begin{equation}
\frac{\partial}{\partial t} \alpha(x,t) =  d_0 \frac{\partial^{2}}{\partial \xi^{2}} 
\alpha(x,t) - \frac{\omega} {\eta} \alpha(x,t) \beta(x,t) \gamma(x,t),
\label{eq:e5}
\end{equation}                             
\\ 
where the perturbation parameter  $d_0$, $\eta$, and $\omega$ are denoted by 
$d_0 ={(D_A / D_B D_C )}^{1/2}$, $ \eta = A_0$,                
and $\omega=K/{(D_A D_B D_C )}^{1/2}$.   
In the perturbation parameters $D_B$ and $D_C$ are the diffusion constants 
for two - species $B$ and $C$. 
For early - time regime, a perturbation theory can be developed to calculate on the decay process 
of three - species reactants , 
provided that the reactive effect is small compared to the diffusion effect.   
For the sake of concreteness,
if it is assumed to be extremely small (i.e. $\omega \ll 1$) for the perturbation coefficient,
the dimensionless particle densities are expanded in series of powers of $\omega$ as        
$\alpha=\sum_{n=1}^{\infty} \alpha_{n} {\omega}^{n}$, $\beta=\sum_{n=1}^{\infty} 
\beta_{n} {\omega}^{n}$,           
and $\gamma=\sum_{n=1}^{\infty} \gamma_{n} {\omega}^{n}$. 
The solutions for $ \alpha(\xi,\tau), \beta(\xi,\tau),$ and $\gamma(\xi,\tau)$ are
obtained that 
\\
\begin{equation}
\alpha(\xi,\tau) = \phi(\frac{\xi}{2{(d_0 \tau)}^{1/2}}),
\beta(\xi,\tau) = \phi(\frac{\xi}{2{(d_0 \tau)}^{1/2}}),
\label{eq:g7}
\end{equation}                             
\\ 
%
%
%
%
%
and
\begin{equation}
\gamma(\xi,\tau) = 1-\phi[\frac{\xi}{2}{(\frac{d_0}{\tau})}^{1/2}],
\label{eq:i9}
\end{equation}                             
\\
where
the dimensionless time $\tau$ and space variables $\xi$ 
are, respectively, given by
$\tau=tA_0 B_0 C_0 (D_A D_B D_C )^{1/2}$, $\xi=x(A_0 B_0 C_0 )^{1/2}$,
and $\phi(x) = \frac{1}{\sqrt{\pi}} \int^{x}_{-\infty} dxe^{-x^{2}}$.\\
\indent
In general, the global reaction rate $R(t)$ also can be defined by
$R(t)=\int^{\infty}_{-\infty}dxR(x,t)$,
where the local reaction rate  is the statistical quantity given by $R(x,t)=K
\alpha(x,t) \beta(x,t) \gamma(x,t)$. 
For simplicity,
we can get the global reaction rate 
from lowest order of perturbation terms. Hence, this  is
calculated as
\\
\begin{equation}
\begin{array}{rcl}
R(\tau) & = & A_0 B_0 C_0 (D_A D_B D_C )^{1/2} \omega(R_0 + \omega R_1 + ... ) \\
              & \sim  & {\tau}^{1/2}\\
\end{array}
\label{eq:l12}
\end{equation}                             
\\   
in early - time regime. \\
\indent
In order to confirm 
numerically the analytical result of the global reaction rate, 
we concentrate mainly on the Monte Carlo simulation in 
reaction - diffusion process of $A + B + C \rightarrow D $.
It is supposed that three - species reactants are 
distributed randomly on two - dimensional square lattice with the periodic boundary conditions.
We also assume that the intermediate process existing concurrently 
the combined two - species reactants can be formed, 
when two - species reactants meet each other on the same lattice points.
If two - species reactants meet the third reactant,
these reactants  react and leave immediately on lattice point.
In this reaction - diffusion process, we only restrict ourselves to the case that 
the diffusion constant takes the same value  for each reactant. \\
%
\indent
In our reaction - diffusion system, the respective particle density of $30\%$ for $A$, $B$, 
and $C$
is distributed randomly on a square lattice having $200 \times 200$ lattice points.
When our simulations are performed on $2\times10^2$ realizations, our  reaction constant  
is $1/1500$.
As shown in Fig.$1$, we can directly observe the crossover for the global reaction rate
from our simulation result. 
Therefore, it is found numerically from Fig.$2$ that
the scaling exponent for the slope before the crossover is $0.49$ that is close nearly
to $0.5$ , the result of eq.(~\ref{eq:l12}) derived from our three - species process.  
The scaling exponent for the slope after the crossover also is $-0.54$,
as shown in Fig.$3$.\\
\indent
In conclusion, 
we have analytically derived 
the global reaction rate before the crossover
in reaction - diffusion system of the form $A + B + C \rightarrow D $.
It is really found from our simulation result that
the global reaction rate is approximately propertional  
to $t^{1/2}$ at early - time regime,
and to $t^{-1/2}$ at long - time regime by scaling arguments.
However, future work is in progress to extend to the systematic method of
renormalization field theory for multi - components reaction - diffusion process,
and  we also will attempt to investigate extensively on fractal lattice models.$^{11)}$\\  
\indent
This work is supported in part by the academic research fund of Ministry of Education of Korea.\\
%
%
%
%

%
%
%
%
\newpage
\section*{Figure Captions}
\vspace{2.0cm}
Fig.1. \\
\\
The global reaction rate $R(t)$ verse time $t$ for the reaction - diffusion process
of the form $A+B+C\rightarrow D$. 
Our numerical simulation is done for
$2\times10^2$ configurations on $200\times200$  square lattice 
with the periodic boundary conditions.  \\[0.5cm]
\\
\\
\\
%
%
%
Fig.2\\
\\
Plot of $lnR(t)$ verse $lnt$ before the crossover.
The slope of solid line is $0.49$.\\[0.5cm]
\\
\\
\\
Fig.3\\
\\
Plot of $lnR(t)$ verse $lnt$ after the crossover. The scaling exponent
we obtained is approximately $-0.54$.\\[0.5cm]
\\
\\
\\
%
%
%
%
%
%
%
\end{document}